\documentclass[12pt]{article}

\usepackage[letterpaper, margin=1.2in, noheadfoot]{geometry}
\usepackage{amsmath}
\usepackage{amssymb}
\usepackage{graphicx}
\usepackage{natbib}
\usepackage{color}

\usepackage{url}

\newcommand{\s}{t}
\newcommand{\bs}{\mathcal{S}}
\newcommand{\bt}{\mathcal{T}}
\newcommand{\g}{g}
\newcommand{\G}{G}
\newcommand{\var}{\sigma}
\newcommand{\pd}{\partial}
\newcommand{\R}{L}
\newcommand{\I}{\alpha}
\newcommand{\p}{p}

\begin{document}
\title{Statistical distributions of sequencing by synthesis 
with probabilistic nucleotide incorporation}
\author{Yong Kong\\
Department of Molecular
Biophysics and Biochemistry\\
W.M. Keck Foundation Biotechnology Resource Laboratory \\
Yale University\\
333 Cedar Street, New Haven, CT 06510\\
email: \texttt{yong.kong@yale.edu} }

\date{}

\maketitle

\newpage

\begin{abstract}
Sequencing by synthesis is used in many
next-generation DNA sequencing technologies.
Some of the technologies, especially those exploring the principle of
single-molecule sequencing,
allow incomplete nucleotide incorporation in each cycle.
We derive statistical distributions
for sequencing by synthesis
by taking into account the possibility
that nucleotide incorporation may not be complete in each flow cycle.
The statistical distributions are expressed
 in terms of nucleotide probabilities
of the target sequences and the nucleotide incorporation probabilities
for each nucleotide.
We give exact distributions both for fixed number of flow cycles
and for fixed sequence length.
Explicit formulas are derived for the mean and variance
of these distributions.
The results are generalizations of our previous work for pyrosequencing.
Incomplete nucleotide incorporation leads to
significant change in the mean and variance of the distributions, 
but still they can be approximated
by normal distributions with the same mean and variance.
The results are also generalized to handle
sequence context dependent incorporation.
The statistical distributions will be useful 
for instrument and software development for sequencing by synthesis
platforms.

\end{abstract}

\newpage

\section{Introduction} \label{S:intro}

The next-generation DNA sequencing technology is changing biological
research in many aspects and opening up new directions of experimental
designs. 
The underlining technology of many of 
the current available platforms 
and those that are still under development  
can be categorized as sequencing by synthesis (SBS).
In SBS, 
nucleotides are added to the reactions repeatedly in a pre-determined 
cyclic manner.
Those nucleotides
that are complementary to the
template sequence will be potentially incorporated
at each step, usually driven by enzymes.
The presence or absence of any signal at each step
reveals the nature of the template being sequenced. 

In some of these SBS technologies, such as the Illumina Genome Analyzer, 
a mixture of four nucleotides is added at each cycle
and the identity of the nucleotide is distinguished by the four different
fluorescent labels.
The nucleotides are modified so that only one base can be incorporated
in each cycle.
In this case, the length of sequencing reads has a simple relation with
the number of sequencing cycles: they are equal to each other.
In some other SBS technologies, such as pyrosequencing,
however, only one kind of nucleotides is added to the reaction in each cycle.
The length of sequencing reads will not have a fixed relation
with the number of sequencing cycles.
Rather, the read length is dependent on the sequence context
and the nucleotide flow order.
For the traditional pyrosequencing, 
the nucleotide incorporation 
is optimized to be as complete as possible in each cycle to avoid errors
caused by dephasing (discussed below).
At each extension cycle, 
ideally the nucleotide should be incorporated $100\%$
if it is complementary to the target sequence
(including the homopolymer regions). 
Analytical statistical distributions 
have been derived for this situation in a previous paper 
\citep{Kong2008}.
In the present paper we deal with statistical distributions
for the case where the nucleotide incorporation
is not achieved $100\%$ at each cycle.
For the emerging single-molecule DNA sequencing technology
\citep{Gupta2008},
sometimes it is desirable to \emph{not} achieve $100\%$ nucleotide incorporation
at each cycle in order to maximize the resolution of homopolymer regions
and increase the accuracy of nucleotide incorporation
\citep{Harris2008}.

In this paper we derive statistical distributions
for SBS with probabilistic, or incomplete,  nucleotide incorporation.
Mathematically these results are generalizations of the results
obtained previously for $100\%$ nucleotide incorporation at each cycle
\citep{Kong2008}.
These distributions are useful in different stages
in the use and development of the next-generation DNA sequencing technology,
such as
instrument development and testing, 
algorithm and software development,
and the everyday machine performance monitoring and trouble-shooting.

The paper is organized as follows.
First in the remaining of this \emph{Introduction} section
 we give a brief description of the single-molecule DNA sequencing technology
(SMS),
for which the theoretical results of present work will find most useful.
We also define the necessary notation here.
The derivation of the main results, 
which are exact under the assumptions of the sequence model,
will be presented in the \emph{Bivariate Generating Functions} section.
After that, we present the explicit 
formulas for the mean and variance of the distributions,
for both fixed number of cycles and fixed sequences length.
After that, we generalize the results to 
the situation where nucleotide incorporation depends on
sequence context.
The main results are summarized in 
Eqs.~\eqref{E:Gs}, \eqref{E:G_sum}, \eqref{E:fixed_length_avg_var}, 
\eqref{E:fixed_cycle_avg_var}), 
and \eqref{E:gfs_g}.


\subsection{Single-molecule DNA sequencing}

Single-molecule DNA sequencing (SMS) technology becomes commercially available 
only in 2008 
\citep{Harris2008}, but major efforts are underway to develop
different kinds
of SMS technologies and it is expected that more SMS platforms
will be available
 in the next few years
\citep{Gupta2008}. Some of these SMS technologies employ
SBS as the underlining principle.

One of the advantages of SMS compared with bulk sequencing technologies
is that, 
unlike other bulk sequencing platforms, 
SMS does not have a clonal amplification step for the target sequences. 
Instead, a single target sequence molecule is used as the template.
This avoids the problems of bias and errors introduced
by PCR in the amplification step.
Another advantage of SMS is that the problem of dephasing associated 
with bulk sequencing can be circumvented.

In the bulk sequencing, a significant copy numbers of the target sequence
have to be present to obtain a detectable signal.
During the sequencing process, 
the synchronization between identical individual templates
will be lost gradually, leading to signal decay and sequencing errors.
To avoid dephasing, the reaction reagents and synthesis chemistry
are usually tuned to drive the enzymatic incorporation to completion
at each cycle.
As a consequence, the misincorporation rate will increase.
For SMS, since each molecule is monitored individually,
the dephasing problem does not exist.
The reaction kinetics can be controlled to adjust the rate of incorporation
to the benefit of sequencing accuracy.
For example, slow reaction kinetics can be used to limit incorporation
to two or three bases
\citep{Harris2008}.
For a homopolymer region \verb+GGG+, 
the bulk sequencing will try to incorporate three \verb+C+'s
in one cycle; 
for SMS, however, zero, one, two, or three \verb+C+'s
can be incorporated in a single synthesis cycle.
This flexibility in 
incorporation rate can be utilized to increase the resolution
of homopolymer region
\citep{Harris2008}.

As a result, for SMS the consecutive zeros in the signal 
(where there is no incorporation) 
is not limited to $3$ as in the pyrosequencing
case.  Instead, if the nucleotide complementary to the template
is not incorporated in the current cycle, it still has chance to be
incorporated in the next cycle, or the next-next cycle,
without detrimental effect on sequencing quality.
The relation between the read length and flow cycle in SMS 
will also be different from that of traditional pyrosequencing.
In the following we will derive an analytical expression for
the distribution, which will depend on nucleotide composition of
the target sequences as well as the incorporation
probability of each nucleotide.

%
%
%


%
%
%
%

\subsection{Notation and definitions}

To avoid the unnecessary specification of the
detailed names of the four kinds of nucleotides, in the following we will
use $a$, $b$, $c$, and $d$ to represent any permutations of 
the usual nucleotides $A$, $C$, $G$, and $T$,
as we did previously
\citep{Kong2008}.
Throughout the paper we assume that the nucleotides in the target
sequence are independent of each other.
Note that this is different from the sequence-context
dependence of \emph{nucleotide incorporation}, which will be addressed
in Section~\ref{S:dependent}.
The probabilities for the four nucleotides in the target sequence
are denoted as 
$p_a$, $p_b$, $p_c$, and $p_d$. 

The nucleotide incorporation probabilities are denoted as
$\I_{ij}$, where $i=a$, $b$, $c$, and $d$ stands for the kind of nucleotides,
$j=0$, $1$, $2$, $\dots$ denotes the cycle number, with the
current cycle as $0$. 
For example, when it is complementary to the template,
if the chance for nucleotide $b$ 
to be incorporated in the current cycle, 
the next cycle, and the next-next cycle
is
$1/3$, $1/2$, and $1/6$, respectively, 
then 
$\I_{b0} = 1/3$,
$\I_{b1} = 1/2$,
$\I_{b2} = 1/6$,
and $\I_{bj} = 0$ for $j > 2$.
Obviously, the pyrosequencing discussed previously is a special
case in this notation with $\I_{i0} = 1$
and $\I_{ij} = 0$ for $j > 0$.
By definition $\sum_{j=0}^\infty \I_{ij} = 1$.
The generating functions (GFs) of the nucleotide
incorporation probabilities $\g_i(x)$ ($i=a$, $b$, $c$, and $d$)
and their associated elementary symmetric functions (ESFs) $\s_i(x)$
($i=1$, $2$, $3$, and $4$)
will be defined in the next section (Eqs.~\eqref{E:gfni} and
~\eqref{E:esfg}).
The $\I_{ij}$ will be generalized to be sequence-context dependent
in  Section~\ref{S:dependent}.

We also use the same definition of nucleotide \emph{flow cycle number}
as before \citep[Table 1]{Kong2008},
which is defined as 
the ``quad cycle'' of successive four nucleotides \{$abcd$\}.
The cycle number is denoted as $f$ in the following.
We will use $n$ for the length of a sequence. 
The relation between nucleotide flow and cycle number
is illustrated in Table~\ref{T:cycle_number}.


\begin{table}
\caption{
The relation between nucleotide flow and cycle number.
}
\label{T:cycle_number}
\begin{tabular}{lc | cccc | cccc | cccc | cccc | c}
cycle number &
$\cdots$      &
\multicolumn{4}{|c|}{$f$} 
&
\multicolumn{4}{|c|}{$f+1$} 
&
\multicolumn{4}{|c|}{$f+2$} 
&
\multicolumn{4}{|c|}{$f+3$} 
&
$\cdots$ 
\\ 
\hline
nucleotide flow &
$\cdots$ &
a & b & c & d & 
a & b & c & d &
a & b & c & d & 
a & b & c & d &
$\cdots$\\
\end{tabular}
\end{table}

To extract coefficients from the expansion of GFs,
we use the notation
$[x^n]f(x)$ to denote the coefficient of $x^n$ in the series of
$f(x)$ in powers of $x$.
Similarly, we use $[x^n y^m]f(x,y)$ to denote the 
coefficient of $x^n y^m$ in the bivariate $f(x, y)$.

\section{Bivariate Generating Functions} \label{S:exact}
In this section we first set up a detailed example
of the recurrence relations between probabilities 
of sequence length and flow cycle number,
and then establish a set of equations for the general case.
The set of equations of probabilities cannot be solved
in closed forms.
However, if we transform the equations of probabilities
into their corresponding generating functions (GFs),
then these GFs can be easily obtained.

\subsection{Recurrences}
Let $\R_i(f, n)$, $i=a$, $b$, $c$, and $d$ denote 
the probability (up to a normalization
factor, see below) of sequences with
a length of $n$ that is synthesized in $f$ flow cycles
with the last nucleotide  being $i$.
First let's look at one example.
In this example,
we assume that 
for each nucleotide only the first three incorporation probabilities
are non-zero, i.e., $\I_{ij} = 0$ for
$j > 2$.
The following four recurrence relations can be established.
It might be helpful to refer back to Table~\ref{T:cycle_number}
for the understanding of these recurrences: 
\begin{subequations}
\label{E:recur}
\begin{align}
\label{E:recur_a}
 \R_a(f+3, n+1) &= 
 \p_a 
 \big[
       \R_a(f+3, n) \I_{a0}
     + \R_a(f+2, n) \I_{a1}
     + \R_a(f+1, n) \I_{a2} 
\notag  \\
   & + \R_b(f+2, n) \I_{a0}
     + \R_b(f+1, n) \I_{a1}
     + \R_b(f, n)   \I_{a2}
\notag  \\
   & + \R_c(f+2, n) \I_{a0}
     + \R_c(f+1, n) \I_{a1}
     + \R_c(f, n)   \I_{a2}
\notag  \\
   & + \R_d(f+2, n) \I_{a0}
     + \R_d(f+1, n) \I_{a1}
     + \R_d(f, n)   \I_{a2}
 \big],
\end{align}
\begin{align}
\label{E:recur_b}
 \R_b(f+3, n+1) &= 
 \p_b 
 \big[
       \R_a(f+3, n) \I_{b0}
     + \R_a(f+2, n) \I_{b1}
     + \R_a(f+1, n) \I_{b2} 
\notag  \\
   & + \R_b(f+3, n) \I_{b0}
     + \R_b(f+2, n) \I_{b1}
     + \R_b(f+1, n) \I_{b2}
\notag  \\
   & + \R_c(f+2, n) \I_{b0}
     + \R_c(f+1, n) \I_{b1}
     + \R_c(f, n)   \I_{b2}
\notag  \\
   & + \R_d(f+2, n) \I_{b0}
     + \R_d(f+1, n) \I_{b1}
     + \R_d(f, n)   \I_{b2}
 \big],
\end{align}
\begin{align}
\label{E:recur_c}
 \R_c(f+3, n+1) &= 
 \p_c 
 \big[
       \R_a(f+3, n) \I_{c0}
     + \R_a(f+2, n) \I_{c1}
     + \R_a(f+1, n) \I_{c2} 
\notag  \\
   & + \R_b(f+3, n) \I_{c0}
     + \R_b(f+2, n) \I_{c1}
     + \R_b(f+1, n) \I_{c2}
\notag  \\
   & + \R_c(f+3, n) \I_{c0}
     + \R_c(f+2, n) \I_{c1}
     + \R_c(f+1, n) \I_{c2}
\notag  \\
   & + \R_d(f+2, n) \I_{c0}
     + \R_d(f+1, n) \I_{c1}
     + \R_d(f, n)   \I_{c2}
 \big],
\end{align}
\begin{align}
\label{E:recur_d}
 \R_d(f+2, n+1) &= 
 \p_d 
 \big[
       \R_a(f+2, n) \I_{d0}
     + \R_a(f+1, n) \I_{d1}
     + \R_a(f  , n) \I_{d2} 
\notag  \\
   & + \R_b(f+2, n) \I_{d0}
     + \R_b(f+1, n) \I_{d1}
     + \R_b(f  , n) \I_{d2}
\notag  \\
   & + \R_c(f+2, n) \I_{d0}
     + \R_c(f+1, n) \I_{d1}
     + \R_c(f  , n) \I_{d2}
\notag  \\
   & + \R_d(f+2, n) \I_{d0}
     + \R_d(f+1, n) \I_{d1}
     + \R_d(f, n)   \I_{d2}
 \big].
\end{align}
\end{subequations}

In general, for arbitrary nucleotide incorporation probabilities
$\I_{ij}$, we can write
down the following four recurrences: 
\begin{subequations}
\label{E:recur_g}
\begin{multline}
 \R_a(f, n) = \p_a \Big \{ 
  \sum_{j=0}^f     \R_a(f-j, n-1)   \I_{aj} + 
  \sum_{j=0}^{f-1} \R_b(f-j-1, n-1) \I_{aj} \\
  + 
  \sum_{j=0}^{f-1} \R_c(f-j-1, n-1) \I_{aj} + 
  \sum_{j=0}^{f-1} \R_d(f-j-1, n-1) \I_{aj}
 \Big \} ,
\end{multline}
\begin{multline}
 \R_b(f, n) = \p_b \Big \{ 
  \sum_{j=0}^f     \R_a(f-j, n-1)   \I_{bj} + 
  \sum_{j=0}^f     \R_b(f-j, n-1)   \I_{bj} \\
  + 
  \sum_{j=0}^{f-1} \R_c(f-j-1, n-1) \I_{bj} + 
  \sum_{j=0}^{f-1} \R_d(f-j-1, n-1) \I_{bj}
 \Big \} ,
\end{multline}
\begin{multline}
 \R_c(f, n) = \p_c \Big \{ 
  \sum_{j=0}^f     \R_a(f-j, n-1)   \I_{cj} + 
  \sum_{j=0}^f     \R_b(f-j, n-1)   \I_{cj} \\
  + 
  \sum_{j=0}^f     \R_c(f-j, n-1)   \I_{cj} + 
  \sum_{j=0}^{f-1} \R_d(f-j-1, n-1) \I_{cj}
 \Big \} ,
\end{multline}
\begin{multline}
 \R_d(f, n) = \p_d \Big \{ 
  \sum_{j=0}^f     \R_a(f-j, n-1)   \I_{dj} + 
  \sum_{j=0}^f     \R_b(f-j, n-1)   \I_{dj} \\
  + 
  \sum_{j=0}^f     \R_c(f-j, n-1)   \I_{dj} + 
  \sum_{j=0}^f     \R_d(f-j, n-1) \I_{dj} 
 \Big \} .
\end{multline}
\end{subequations}

The recurrences cannot be solved in closed forms.
However, if we transform these recurrences into their corresponding
GFs, then these GFs can be solved in compact forms.
The bivariate GFs of $L_i(f, n)$ are defined as
\begin{equation}
 \G_i (x, y) 
 = \sum_{n=1}^\infty \sum_{f=1}^\infty \R_i(f, n) x^f y^n,
 \qquad i = a, b, c, d.  
\end{equation}
We also define the GFs for nucleotide incorporation probabilities as
\begin{equation} \label{E:gfni}
 \g_i(x) = \p_i \sum_{j=0}^\infty \I_{ij} x^j
 \qquad i = a, b, c, d.  
\end{equation}
Since $\sum_{j=0}^\infty \I_{ij} = 1$,
we have
\[
 \g_i(1) = \p_i
 \qquad i = a, b, c, d.  
\]

To transform the
system of recurrence equations in Eq.~\eqref{E:recur_g} to a system of 
equations of GFs, we need to use 
the following identities:
\[
 \sum_{f=1}^\infty \sum_{n=1}^\infty \sum_{j=0}^{f-1} 
 \R_i(f-j, n-1) \I_{ij} x^f y^n
= y \left[ \G_i (x, y) + \G_{i0}(x) \right ]  \g_i(x),
 \qquad i = a, b, c, d  
\]
and
\[
 \sum_{f=1}^\infty \sum_{n=1}^\infty \sum_{j=0}^{f-2} 
 \R_i(f-j-1, n-1) \I_{ij} x^f y^n
= x y \left[ \G_i (x, y) + \G_{i0}(x) \right ]  \g_i(x),
 \qquad i = a, b, c, d  
\]
where
\[
 \G_{i0}(x) = \sum_{f=1}^{\infty} \R_i(f, 0) x^f, 
 \qquad i = a, b, c, d.  
\]
By definition, we have 
\begin{align*}
 \G_{a0}(x) &= x, \\
 \G_{b0}(x) &= 0, \\
 \G_{c0}(x) &= 0, \\
 \G_{d0}(x) &= 0. 
\end{align*}
The system of equations of GFs after the transform is:
\begin{subequations}
\label{E:gfs}
\begin{align}
 \G_a &=  \Big[ \G_a + x  + x \G_b  + x \G_c  + x \G_d \Big ] y \g_a, \\
 \G_b &=  \Big[ \G_a + x  +   \G_b  + x \G_c  + x \G_d \Big ] y \g_b, \\
 \G_c &=  \Big[ \G_a + x  +   \G_b  +   \G_c  + x \G_d \Big ] y \g_c, \\
 \G_d &=  \Big[ \G_a + x  +   \G_b  +   \G_c  +   \G_d \Big ] y \g_d
\end{align}
\end{subequations}
which can be solved as
\begin{subequations}
\label{E:Gs}
\begin{align}
 \G_a (x, y) &= \frac{\g_a x y} {H} F    ,        \label{E:G_a}\\
 \G_b (x, y) &= \frac{\g_b x y }{H}  
 \left[ 1 - (\g_c + \g_d) (1-x) y  
   + \g_c \g_d (1-x)^2 y^2  
 \right]                                 ,        \label{E:G_b}\\
 \G_c (x, y) &= \frac{\g_c x y}{H} 
 \left[ 1 -   \g_d (1-x) y \right]       ,        \label{E:G_c}\\
 \G_d (x, y) &= \frac{\g_d x y} {H}      ,        \label{E:G_d}
\end{align}
\end{subequations}
where
\begin{equation}
  \label{E:H}
 H = 1 - \s_1 y + \s_2 (1-x) y^2 - \s_3 (1-x)^2 y^3 + \s_4 (1-x)^3 y^4 ,
\end{equation}
and
\begin{multline*}
 F = 
   [ 1 - (\g_b + \g_c + \g_d) (1-x) y  \\
     + (\g_b \g_c + \g_b \g_d + \g_c \g_d) (1-x)^2 y^2 
     - \g_b \g_c \g_d (1-x)^3 y^3
   ] .
\end{multline*}
Here we use \emph{elementary symmetric functions} (ESFs) $\s_i(x)$ of 
the nucleotide incorporation probabilities GFs $\g_i(x)$ to put the
solutions of $\G_i(x, y)$ into a more compact form.
These ESFs $\s_i(x)$ of four variables are defined as:
\begin{align} \label{E:esfg}
  \s_1 (x) &= \g_a + \g_b + \g_c + \g_d ,                       \notag \\
  \s_2 (x) &= \g_a \g_b + \g_a \g_c +  \g_a \g_d + 
  \g_b \g_c + \g_b \g_d + \g_c \g_d ,\notag\\ 
  \s_3 (x) &= \g_a \g_b \g_c +  \g_a \g_b \g_d + \g_a \g_c \g_d + 
  \g_b \g_c \g_d ,    \notag\\
  \s_4 (x) &= \g_a \g_b \g_c \g_d .                                  
\end{align}
If we put $x=1$ into $\s_i(x)$, we get back the ESFs $s_i$
of nucleotide probabilities $\p_i$ in the target sequences,
which we used in the previous work
\citep{Kong2008}:
\begin{align} \label{E:esf}
  s_1 &= \s_1 (1) = p_a + p_b + p_c + p_d = 1,                     \notag \\
  s_2 &= \s_2 (1) = p_a p_b + p_a p_c +  p_a p_d + p_b p_c 
  + p_b p_d + p_c p_d ,\notag\\ 
  s_3 &= \s_3 (1) = p_a p_b p_c +  
  p_a p_b p_d + p_a p_c p_d + p_b p_c p_d ,    \notag\\
  s_4 &= \s_4 (1) = p_a p_b p_c p_d .                                           
\end{align}
For complete nucleotide incorporation, $\g_i(x) = \p_i$. In this case
we have $\s_i(x) = s_i$ as a constant, instead of a function of $x$.

We see from the expressions of $G_i(x,y)$ in Eq.~\eqref{E:Gs} that
they are identical in \emph{forms} to the solutions in the traditional
pyrosequencing where $100\%$ incorporation is assumed
\citep{Kong2008}, with ESFs $\s_i(x)$ of
the nucleotide incorporation probabilities GFs $\g_j(x)$
replacing the ESFs $s_i$ of nucleotide probabilities $\p_j$, 
$i=1$, $2$, $3$, $4$ and $j=a$, $b$, $c$, $d$.

From the expressions of the GFs in Eq.~\eqref{E:Gs} 
we can see that they are not symmetric
with respect to the nucleotide incorporation probability GFs 
$\g_i(x)$.
If we only consider the nucleotide flows  
that end up in the same ``quad cycle'' (see Table~\ref{T:cycle_number}),
then we can add the four GFs  $G_i(x, y)$ together
to obtain a symmetric GF
\begin{align} \label{E:G_sum}
G (x, y) &= \sum_{n=1}^\infty \sum_{f=1}^\infty \R(f, n) x^f y^n
          = G_a + G_b + G_c + G_d \notag \\
  &= \frac{xy}{H} [ \s_1 - \s_2(1-x)y + \s_3(1-x)^2 y^2 - \s_4(1-x)^3 y^3] .
\end{align}
Here $\R(f, n)$ is the (unnormalized) probability
of a sequence with
a length of $n$ which can be synthesized in $f$ flow cycles.

\subsection{Normalization factors} \label{SS:norm}
To treat $\R_i(f, n)$ as real probabilities, they have to be normalized.
As we did in the previous work, 
these normalization factors can be obtained
by setting $x=1$ and $y=1$ in the GFs in 
Eq.~\eqref{E:Gs} or Eq.~\eqref{E:G_sum}.
If we set $x=1$, then we obtained the normalization factors
for $L_a(f, n)$
when the sequence length is fixed at $n$.
From Eq.~\eqref{E:Gs} we get
\[
 \G_i(1, y) = \frac{p_i y}{1 - y},
\]
from which we obtain the normalization factors $u_i$ for $\R_i(f, n)$ as
\begin{equation} \label{E:nf_ui}
 u_i = \sum_{f=1}^\infty \R_i(f, n) = [y^n]\G_i(1, y) = \p_i
\end{equation}
The normalization factor for $\R(f, n) = \sum_i \R_i(f, n)$
is simply the sum of $u_i$:
\[
 u = u_a + u_b + u_c + u_d = 1 .
\]

For fixed cycle, by setting $y=1$, the denominator of $\G_i(x, 1)$ becomes
\[
 1 - \s_1 + \s_2 (1-x) - \s_3(1-x)^2 + \s_4 (1-x)^3 .
\]
Since $\s_1(1) = 1$,
we see that $x=1$ is a root of the denominator of $\G_i(x, 1)$,
 and it is the dominant part 
in the expansion of $\G_i(x, 1)$:
\[
  \G_i(x, 1) = \frac{\beta_{i1}}{1-x} + \cdots.
\]
The coefficients of the expansion can be evaluated as
\[
 \beta_{i1} = (1-x) \G_i(x, 1) |_{x=1} = \frac{\p_i}{s_2 + \s_1'(1)}, 
\]%
%
from which come the normalization factors
for $L_i(f, n)$
when the number of flow cycles is fixed at $f$:
\begin{equation} \label{E:nf_vi}
v_i = \sum_{n=1}^\infty L_a(f, n) = [x^f]\G_i(x, 1) 
  \approx \frac{p_i}{s_2 + \s_1'(1)} .
\end{equation}
Here $\s_1'(x) = \pd \s_1(x) / \pd x $ and
\[
 \s_1'(1) = \frac{\pd \s_1(x) } {\pd x} \bigg |_{x=1}
\]
 stands for
the value of the first derivative of $\s_1(x)$ evaluated at $x=1$.
 
The normalization factor for $\R(f, n) = \sum_i \R_i(f, n)$
when the number of flow cycles is fixed at $f$ is
\begin{equation} \label{E:nf_v}
 v = v_a + v_b + v_c + v_d \approx \frac{1}{s_2 + \s_1'(1)} .
\end{equation}
As we will see below, $v$ is an important part in the expression
of mean and variance for $\R_i(f, n)$.
When the normalization factor $v$ is compared with 
that of the complete incorporation case
\citep{Kong2008}, the only difference is the extra term of $\s_1'(1)$.

\subsection{Mean and variance}
The availability of GFs makes it easy to derive the mean and variance
for the distributions of $\R_i(f, n)$.
When the sequence length is fixed,
the mean and variance are given by
\begin{subequations}
 \label{E:avg_var_fixed_length}
\begin{align}
 \bar{f} (n)   &= \frac{1}{u}  [y^n]\frac{\pd G(x, y) } 
     {\pd x} \Big |_{x=1}   , 
 \label{E:avg_fixed_length}\\
 \var^2_f (n)  &= \frac{1}{u}  [y^n]\frac{\pd^2 G(x, y) } 
     {\pd x^2} \Big |_{x=1} 
 + \bar{f}(n) - \bar{f}^2(n) . 
\label{E:var_fixed_length}
\end{align}
\end{subequations}
Similar formulas apply to the individual nucleotide flow GFs $G_a$, $G_b$, 
etc,
with their corresponding normalization factors $u_i$
as shown in Eq.~\eqref{E:nf_ui}.

When the number of flow cycles is fixed,
the mean and variance are given by
\begin{subequations}
 \label{E:avg_var_fixed_cycle}
\begin{align}
 \bar{n} (f)  &= \frac{1}{v} [x^f]\frac{\pd G(x, y) } {\pd y} \Big |_{y=1}  ,  
 \label{E:avg_fixed_cycle}\\
 \var^2_n (f)  &= \frac{1}{v} [x^f]\frac{\pd^2 G(x, y)} {\pd y^2} \Big |_{y=1} 
 + \bar{n}(f) - \bar{n}^2(f) .  
\label{E:var_fixed_cycle}
\end{align}
\end{subequations}
Similar formulas apply to the individual nucleotide flow GFs,
with their corresponding normalization factors $v_i$ 
shown in Eq.~\eqref{E:nf_vi}.

\section{Distributions at fixed sequence length and 
fixed number of flow cycles}

With the help of the expression of GF $G(x, y)$ in Eq.~\eqref{E:G_sum},
we can calculate \emph{exact} distribution of $\R(f, n)$
at a fixed sequence length $n$ or a fixed number of cycles $f$,
by expanding $G(x, y)$ and extracting $x^f$ or $y^n$, respectively. 
By using formulas Eqs.~\eqref{E:avg_var_fixed_length} and 
~\eqref{E:avg_var_fixed_cycle}
discussed above,
we can calculate the mean and variance of these distributions.
We will discuss the two cases separately in the following.

\subsection{Fixed sequence length: distribution of flow cycles} 
\label{ss:fixed_length}

When the length the target sequences is fixed at $n$,
the mean ${\bar f} (n)$ 
and variance $\var_f^2 (n)$ of the number of flow cycles $f$ that is needed
to determine the sequences
are calculated from Eq.~\eqref{E:avg_var_fixed_length} as
\begin{subequations} 
\label{E:fixed_length_avg_var}
\begin{equation} 
 {\bar f} (n) = [s_2 + \s_1'] n - s_2 + 1 
\label{E:fixed_length_avg} ,\\ 
\end{equation} 
\begin{multline}
 \var_f^2 (n) = [s_2 - 3 s_2^2 + 2 s_3 
     + \s_1' + 2 \s_2' + \s_1'' - 4 \s_1' s_2 - \s_1'^2] n \\
 + [ 5 s_2^2 - s_2 - 4 s_3 -2 \s_2' + 4 \s_1' s_2 ] .
 \label{E:fixed_length_var}
\end{multline}
\end{subequations} 
Here and in the following we use the abbreviations such as
$\s_1'$ to denote $\s_1'(1)$, $\s_1''$ as 
$\s_1''(1) = \pd^2 \s_1(x) /\pd x^2 |_{x=1}$, 
etc.
They are the derivatives of $\s_i(x)$ with respect to $x$ 
evaluated at the value of $x=1$.

From Eqs. \eqref{E:fixed_length_avg} and \eqref{E:fixed_length_var} we can see
that both the mean ${\bar f} (n)$ and the variance $\var_f^2 (n)$
of flow cycles
increase linearly with the sequence length $n$.
Compared with the complete incorporation situation
\citep{Kong2008}, the difference is the extra terms of the derivatives
of $\s_i(x)$ evaluated at the value of $x=1$.
When the nucleotide incorporation is complete at each cycle, 
$\s_i(x) = \p_i$ so all these derivatives
vanish, leading back to the original results of pyrosequencing
\citep{Kong2008}.

\begin{figure}
  \centering
  \includegraphics[angle=270,width=\columnwidth]{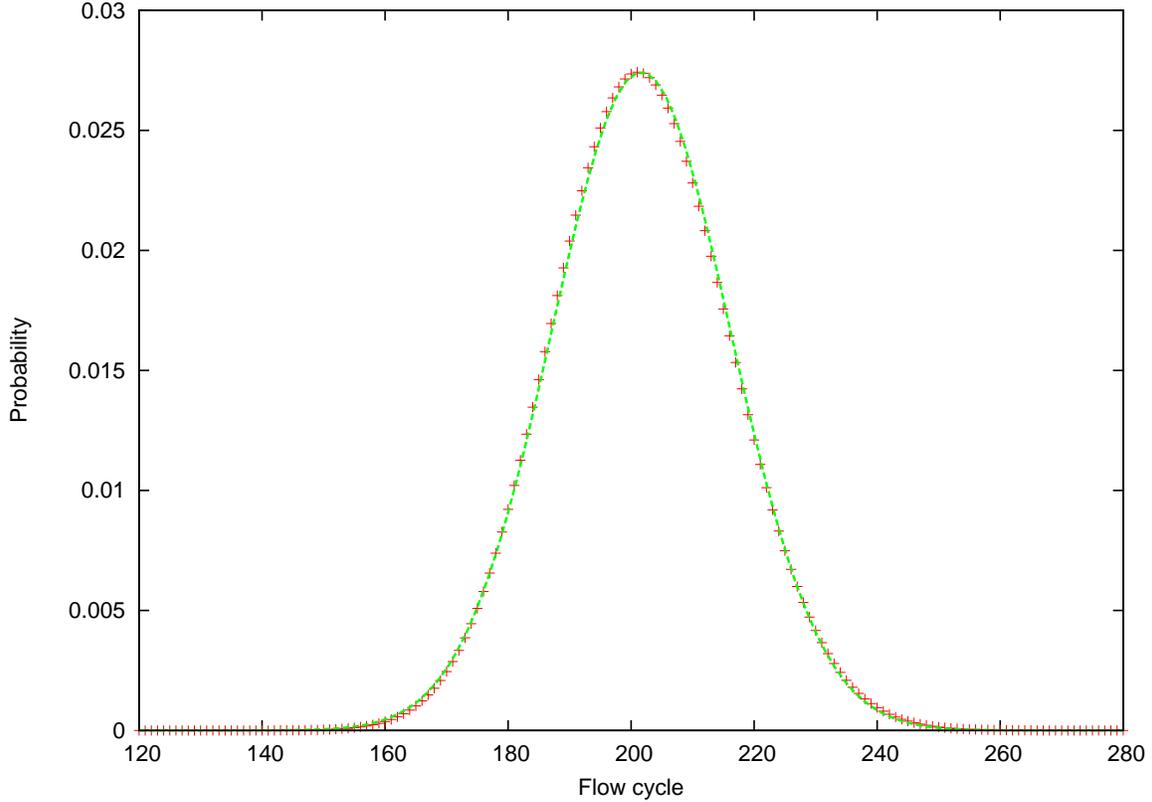}
  \caption{
    The distribution of flow cycles for a fixed sequence
    length of $n=100$ base pairs.
    The nucleotide composition probabilities used here are
    $\p_a=3/10=0.3$, $\p_b=1/5=0.2$, $\p_c = 1/5=0.2$,
    and $\p_d=3/10=0.3$.
    The non-zero nucleotide incorporation probabilities are
    $\I_{aj} = [1/10, 1/5, 2/5, 3/10]$,
    $\I_{bj} = [3/10, 1/5, 1/10, 1/10, 1/10, 1/10, 1/10]$,
    $\I_{cj} = [3/10, 3/10, 3/10, 1/10]$,
    and 
    $\I_{dj} = [2/5, 1/5, 1/5, 1/10, 1/10]$.
    The exact distribution is plotted as '+' and 
    is calculated from Eq.~\eqref{E:G_sum}.
    The continuous curve is 
    the normal distribution $N({\bar f} (n), \var_f^2 (n))$
    of the same mean and variance as those of the exact distribution,
    where ${\bar f} (n)$ and $\var_f^2 (n)$ are calculated
    from Eqs.~\eqref{E:fixed_length_avg} and \eqref{E:fixed_length_var}.
    The normal distribution shown here is
    $N(201.63, 211.5197)$.
    \label{F:fixed_sequence_length}} 
\end{figure}

In Figure~\ref{F:fixed_sequence_length}
the exact distributions of flow cycles are shown for a fixed sequence
length of $n=100$ base pairs with an artificial example of
unequal nucleotide composition probabilities $\p_i$ and
 nucleotide incorporation probabilities $\I_{ij}$.
The nucleotide probabilities used here are
$\p_a=3/10=0.3$, $\p_b=1/5=0.2$, $\p_c = 1/5=0.2$,
and $\p_d=3/10=0.3$.
The non-zero nucleotide incorporation probabilities $\I_{ij}$ are 
(from $j=0$ and up)
$\I_{aj} = [1/10, 1/5, 2/5, 3/10]$,
$\I_{bj} = [3/10, 1/5, 1/10, 1/10, 1/10, 1/10, 1/10]$,
$\I_{cj} = [3/10, 3/10, 3/10, 1/10]$,
and 
$\I_{dj} = [2/5, 1/5, 1/5, 1/10, 1/10]$.
To keep numerical precision,
we used exact calculation throughout by using
either integers or exact fractions 
for all the coefficients in the expansion of the GFs,
as we did in the previous work
\citep{Kong2008}.
The expansion was done using PARI/GP, 
a computer algebra system~\citep{PARI2008}.

Also shown in Figure~\ref{F:fixed_sequence_length} 
in continuous curve
is the 
normal distribution $N({\bar f} (n), \var_f^2 (n))$,
the mean ${\bar f} (n)$ and variance $\var_f^2 (n)$ of which are calculated
from Eqs. \eqref{E:fixed_length_avg} and \eqref{E:fixed_length_var}.
It is evident that,
as in the complete nucleotide incorporation situation,
the exact distributions can be approximated
quite accurately by normal distributions
with the same mean and variance.
For our example here,
the normal distribution is 
$N(201.63, 211.5197)$.
Compared with the complete nucleotide incorporation case,
the incomplete incorporation significantly increases the number of flow
cycles needed to cover the sequences of the same length.
At the same time, the variance also increases significantly.
For complete nucleotide incorporation 
with the same nucleotide composition probabilities 
$\p_i$ in the target sequences
and at the same fixed sequence length $n=100$, 
the mean and variance of flow cycle are $37.63$ and $8.0045$, respectively
\citep{Kong2008}.


\subsection{Fixed flow cycle: distribution of sequence length} 
\label{ss:fixed_cycle}

When the number of flow cycles $f$ is fixed,
the mean ${\bar n}(f)$ 
and variance $\var_n^2 (f)$ of the length of the  sequences  that
can be determined by these flow cycles
are calculated by Eqs~\eqref{E:avg_fixed_cycle} and 
\eqref{E:var_fixed_cycle} as:
\begin{subequations}
\label{E:fixed_cycle_avg_var}
\begin{equation} 
 {\bar n} (f) \approx v f - v^2 [2 s_2^2 - 2 s_3 
    + 3 \s_1' s_2 + \s_1'^2 - \s_1'' - 2 \s_2'] ,
  \label{E:fixed_cycle_avg}
\end{equation}
\begin{multline}
 \var_n^2 (f) \approx  v^3  w  f 
 - v^4 \big [2 s_2 (s_2 s_3 + 3  s_4)
       - 2 s_3 \s_1' (3 \s_1' + 2 s_2) 
       + 6 \s_1' s_4
       - 2(2 \s_2' + \s_1'' + 2 s_3)^2 
\\
              + \big (3 \s_2''  + \s_1'''  + 6 \s_3' 
              + \s_1'' (5 s_2 + \s_1')  
              - 2 \s_2' (\s_1' - 3 s_2) \big ) / v 
	      + \s_1' s_2 / v^2 \big ] ,
 \label{E:fixed_cycle_var} 
\end{multline} 
\end{subequations}
where
\[
w = s_2 - 3 s_2^2 + 2 s_3 
     + \s_1' + 2 \s_2' + \s_1'' - 4 \s_1' s_2 - \s_1'^2.
\]
Eqs.~\eqref{E:fixed_cycle_avg} and \eqref{E:fixed_cycle_var} 
shows that both the average sequence length ${\bar n} (f)$ 
and the variance $\var_n^2 (f)$
increase linearly with the number of flow cycle $f$.
As discussed in section~\ref{SS:norm},
the small extra terms are ignored in the above expressions.

\begin{figure} 
  \centering
  \includegraphics[angle=270,width=\columnwidth]{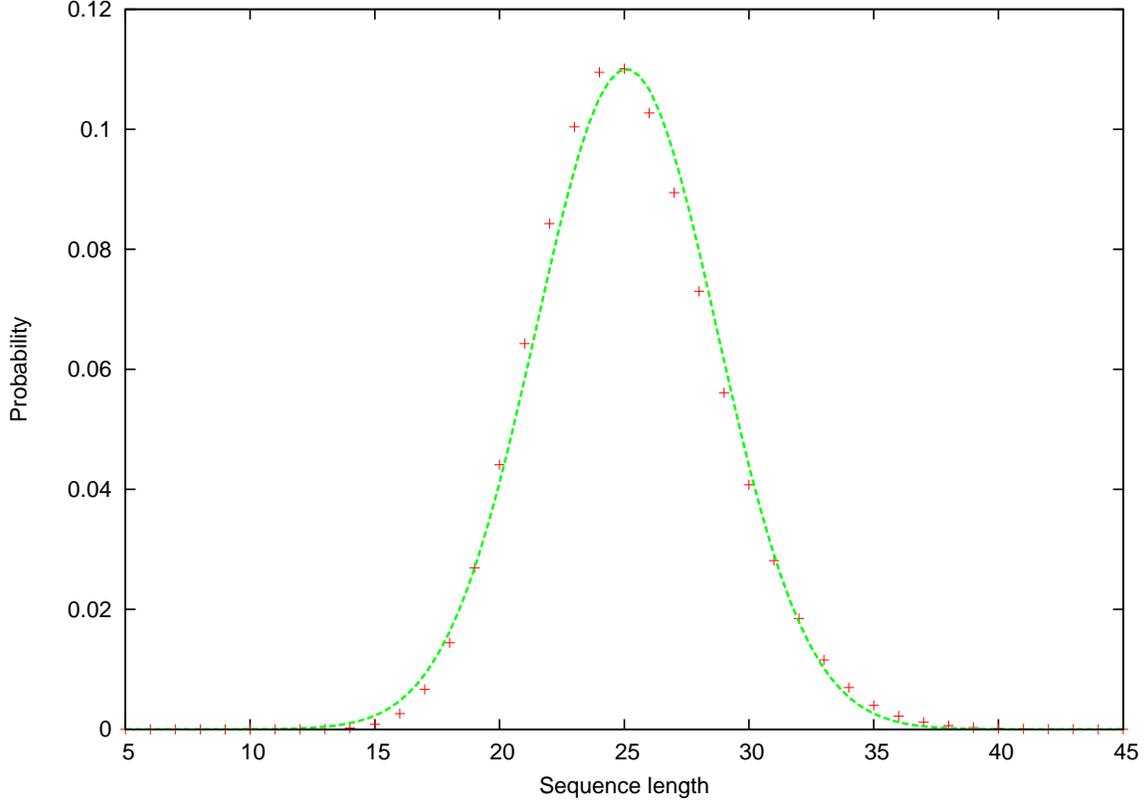}
  \caption{
    The distributions of sequence length in base pairs
    for a fixed number of flow cycles
    $f=50$.
    The nucleotide composition probabilities $\p_i$ 
    and nucleotide incorporation probabilities $\I_{ij}$
    used here are
    the same as in Figure~\ref{F:fixed_sequence_length}.
    The exact distribution is plotted as '+' and 
    is calculated from Eq.~\eqref{E:G_sum}.
    The continuous curve is 
    the normal distribution $N({\bar n} (f), \var_n^2 (f))$
    with the same mean and variance as those of the exact distribution,
    where ${\bar n} (f)$ and $\var_n^2 (f)$ are calculated
    from Eqs.~\eqref{E:fixed_cycle_avg} and \eqref{E:fixed_cycle_var}.
    The normal distribution  shown here is 
    $N(25.0856, 13.1454)$.   
    \label{F:fixed_cycle}} 
\end{figure}

In Figure~\ref{F:fixed_cycle}
the exact distribution of sequence length in base pairs is shown for a fixed
number of flow cycles $f=50$ ($200$ nucleotide flows).
The nucleotide composition probabilities $\p_i$ 
and nucleotide incorporation probabilities $\I_{ij}$
used here are the same as in the
previous section.
These exact distribution is calculated from  Eq.~\eqref{E:G_sum}
in section~\ref{S:exact}.

Also shown in Figure~\ref{F:fixed_cycle} 
in continuous curve
is the 
normal distribution $N({\bar n} (f), \var_n^2 (f))$,
the mean ${\bar n} (f)$ and variance $\var_n^2 (f)$ of which are calculated
from Eqs.~\eqref{E:fixed_cycle_avg} and \eqref{E:fixed_cycle_var}.
Just like the distributions of the number of flow cycles at fixed sequence
length as discussed in the previous section,
the exact distributions of sequence length at a fixed  number of flow cycles
can also be approximated well with normal distributions
with the same mean and variance
as those of the exact distributions.
For the example here, the 
normal distributions is 
$N(25.0856, 13.1454)$.
The introduction of incomplete  nucleotide incorporation
significantly reduces the mean and variance of the read
length of sequences that can be determined at a given number of flow cycles.
For complete  nucleotide incorporation,
with all other parameters being the same as above,
the mean and variance of sequence length for this example would be
$134.0117$ and $78.5114$, respectively
\citep{Kong2008}.

Compared with the fit between the exact and normal
distributions in Figure~\ref{F:fixed_sequence_length},
the curve of normal distribution in Figure~\ref{F:fixed_cycle}
shows some disagreements with the exact distribution,
with the exact distribution having a slightly longer tails 
on the right and 
a slightly shorter tail on the left when compared to the normal distribution.
The discrepancy is similar to that found in the 
complete  nucleotide incorporation situation
\citep[Figure 2]{Kong2008}.


\section{Generalization to sequence context dependent incorporation}
\label{S:dependent}
In the previous discussions we assume that the nucleotide incorporation
does not depend on sequence context.  
This may be only a first order approximation to the real situation
\citep{Harris2008}.
The results in previous sections, however, can be generalized
to take into account sequence context dependent nucleotide
incorporation.
Instead of using $\I_{ij}$, which describes the probability of
nucleotide $i$ being incorporated in the $j$\-th cycle,
we can introduce $\I_{kij}$,
which is the probability of 
nucleotide $i$ being incorporated in the $j$\-th cycle
if the previous incorporated nucleotide is $k$.
Correspondingly,
the nucleotide incorporation GFs will become
\[
 \g_{ki} (x) = \p_i \sum_{j=0}^\infty \I_{kij} x^j,
 \qquad i, k = a, b, c, d.  
\]
Instead of four $\g_i(x)$, we now have $16$ $\g_{ki} (x)$.
The system of equations of GFs in Eq.~\eqref{E:gfs} will become
\begin{subequations}
\label{E:gfs_g}
\begin{align}
 \G_a &=  \Big[ (\G_a + x) \g_{aa}  + x \g_{ba} \G_b  
   + x \g_{ca} \G_c  + x \g_{da}\G_d \Big ] y   , \\
 \G_b &=  \Big[ (\G_a + x) \g_{ab}  +   \g_{bb} \G_b  
   + x \g_{cb} \G_c  + x \g_{db} \G_d \Big ] y  , \\
 \G_c &=  \Big[ (\G_a + x) \g_{ac}  +   \g_{bc} \G_b  
   +   \g_{cc} \G_c  + x \g_{dc} \G_d \Big ] y  , \\
 \G_d &=  \Big[ (\G_a + x) \g_{ad}  +   \g_{bd} \G_b  \
   +   \g_{cd} \G_c  +   \g_{dd} \G_d \Big ] y  .
\end{align}
\end{subequations}
The GFs $\G_i(x, y)$ can be solved in terms of $g_{ki}(x)$.
The solution of $\G = \G_a  + \G_b + \G_c + \G_d$ is in the same form 
as Eq.~\eqref{E:G_sum}:
\[
 \G(x, y) = \frac{x y\left [ 
                \bs_1  - \bs_2 y + \bs_3 y^2 - \bs_4 y^3 \right ]}
           {1 - \bt_1 y + \bt_2 y^2 - \bt_3 y^3 + \bt_4 y^4}
\]
where $\bs_i(x)$ and $\bt_i(x)$ are functions of $\g_{ki}(x)$, and
$\bs_4 = \bt_4$.

For a particular set of nucleotide incorporation probabilities $\I_{kij}$ 
(and hence the $16$ nucleotide incorporation GFs $\g_{ki}$) 
Eq.~\eqref{E:gfs_g}
can be used to solve for $\G_i(x, y)$ for the exact distributions, 
and Eqs.~\eqref{E:avg_var_fixed_length} and ~\eqref{E:avg_var_fixed_cycle} 
can be used to calculate the mean and 
variance for the approximate normal distributions.
The explicit expressions of mean and variance like those 
in Eqs.~\eqref{E:fixed_length_avg_var}
and \eqref{E:fixed_cycle_avg_var}, however, seem difficult to obtain
in compact forms for the sequence context dependent incorporation.

\section{Discussion}
In this paper we derived the statistical distributions 
for SBS with probabilistic nucleotide incorporation.
The solutions are generalizations of the results obtained
previously for pyrosequencing, where nucleotide incorporation
is assumed to be complete for each flow cycle.
Exact distributions can be obtained from the GFs
(Eq.~\eqref{E:Gs}),
and these exact distributions can be approximated by
normal distributions, the mean and various of which
are calculated from the explicit formulas derived from GFs
(Eqs.~\eqref{E:fixed_length_avg_var} and
\eqref{E:fixed_cycle_avg_var}).
The exact distributions can also be obtained
when the nucleotide incorporation is sequence context dependent
(Eq.~\eqref{E:gfs_g}).
 
Probabilistic, or incomplete nucleotide incorporation, 
although a thing to avoid
in traditional bulk sequencing, 
bring benefits for SMS, such as increased accuracy for incorporation
and higher resolution for homopolymer regions.
In SMS each template molecule is monitored individually, 
which makes it possible to bypass the phasing problem
faced by the bulk sequencing technologies.
The potentials of higher throughput and lower cost
make it possible that SMS technologies
will become a major biological and biomedicine research tool
in the near future. 
The statistical distribution derived here
will be useful for instrument and software development
of the next-generation sequencing platforms, including the SMS technologies.

\section*{Acknowledgment}
  This work was supported by
  Yale School of Medicine.

\bibliographystyle{jcbnat}

\end{document}